\documentclass[10pt,letterpaper]{article}
\usepackage[top=0.85in,footskip=0.75in,marginparwidth=2in]{geometry}

\usepackage[utf8x]{inputenc}
\usepackage{textgreek}
\usepackage{cite}
\usepackage{amstext}

\usepackage{nameref,hyperref}

\usepackage[right]{lineno}

\usepackage{microtype}
\DisableLigatures[f]{encoding = *, family = * }

\raggedright
\usepackage{changepage}

\usepackage[aboveskip=1pt,labelfont=bf,labelsep=period,singlelinecheck=off]{caption}

\makeatletter
\renewcommand{\@biblabel}[1]{\quad#1.}
\makeatother

\usepackage{lastpage,fancyhdr,graphicx}
\usepackage{epstopdf}
\fancyheadoffset[L]{2.25in}
\fancyfootoffset[L]{2.25in}

\usepackage{color}

\definecolor{Gray}{gray}{.25}

\usepackage{graphicx}

\usepackage{sidecap}

\usepackage{wrapfig}
\usepackage[pscoord]{eso-pic}
\usepackage[fulladjust]{marginnote}
\reversemarginpar

\newcommand\blfootnote[1]{%
  \begingroup
  \renewcommand\thefootnote{}\footnote{#1}%
  \addtocounter{footnote}{-1}%
  \endgroup
}

\begin{document}

\vspace*{0.35in}

\begin{flushleft}
{\Large
\textbf\newline{Kilohertz binary phase modulator for pulsed laser sources using a digital micromirror device\blfootnote{www.doi.org/10.1364/OL.43.000022 © 2018 Optical Society of America. One print or electronic copy may be made for personal use only. Systematic reproduction and distribution, duplication of any material in this paper for a fee or for commercial purposes, or modifications of the content of this paper are prohibited.}}
}
\newline
\\
Maximilian Hoffmann\textsuperscript{1},
 Ioannis N. Papadopoulos\textsuperscript{1},
 Benjamin Judkewitz\textsuperscript{1,*},
\\
\bigskip
\bf{1} Charit\'{e} Universit\"{a}tsmedizin Berlin and Humboldt University, Einstein Center for Neuroscience, NeuroCure Cluster of Excellence, Berlin, Germany, Charit\'{e}platz 1, 10117 Berlin, Germany\\
\bigskip
* benjamin.judkewitz@charite.de

\end{flushleft}

\begin{abstract}
The controlled modulation of an optical wavefront is required for aberration correction, digital phase conjugation or patterned photostimulation. For most of these applications it is desirable to control the wavefront modulation at the highest rates possible. The digital micromirror device (DMD) presents a cost-effective solution to achieve high-speed modulation and often exceeds the speed of the more conventional liquid crystal spatial light modulator, but is inherently an amplitude modulator. Furthermore, spatial dispersion caused by DMD diffraction complicates its use with pulsed laser sources, such as those used in nonlinear microscopy. Here we introduce a DMD-based optical design that overcomes these limitations and achieves dispersion-free high-speed binary phase modulation. We show that this phase modulation can be used to switch through binary phase patterns at the rate of 20 kHz in two-photon excitation fluorescence applications. 
\end{abstract}

Controlling the amplitude and phase of an optical field, a process known as wavefront shaping, is desirable in many applications. In biomedical imaging it enables deeper imaging into tissue by compensating for scattering or aberrations introduced by refractive index inhomogeneities \cite{Horstmeyer:2015db, Mosk:2012fc}. In addition, several applications of wavefront shaping, such as holographic optogenetic stimulation, dynamic aberration correction during scanning microscopy and fast optical phase conjugation, benefit from precise and fast dynamic control of the amplitude and phase of light. Because of their high switching rates (in the kHz range) and their high pixel count, digital micromirror devices (DMD) can be an attractive and cost-effective solution~\cite{Chandrasekaran:2014ci}.
Consisting of an array of micron-sized mirrors, each of which can be tilted by $\pm 12^\circ$ (mechanical angle) depending on an applied ON or OFF state voltage, DMDs are conventionally used as binary amplitude-only modulators. Since phase modulation results in higher focus enhancement  \cite{Vellekoop:2015gu} in digital optical phase conjugation and generally lower background in holographic applications, a variety of techniques have been explored to turn amplitude modulators, into both binary \cite{Vellekoop:2012cm} and multilevel phase-only modulators \cite{Lee:1978js, Goorden:2014eba, DonConkey:2012vd}. While these previous techniques usually trade off the number of the effective degrees of freedom or diffraction efficiency to achieve phase modulation, they have been successfully applied to wavefront shaping of monochromatic light. In some applications, such as nonlinear microscopy and optogenetic stimulation, which use ultra-short laser pulses as the excitation, broadband laser light (typical bandwidth: ~10 nm) will undergo wavelength dependent diffraction at the surface of the DMD, whose array of tilted mirrors acts as a blazed grating. This results in spatial dispersion and pulse broadening \cite{Gu:2004ef}. This effect can be compensated for by using a diffraction grating as a corrective element, which cancels the spatial dispersion introduced by the DMD \cite{Wang:2014kj, Geng:2017km}, requiring alignment of an additional grating or an identical DMD chip~\cite{Love:2014ba}. However, this amplitude modulation is inherently lossy and the theoretically achievable modulation efficiency is limited by the fact that on average half of the incident light is blocked by DMD mirrors in the OFF position.

Here we introduce a novel method to obtain arbitrary binary phase modulated wavefronts using a DMD, while simultaneously correcting for spatial dispersion introduced by the underlying grating structure without the use of additional diffractive elements. 
This is achieved by splitting the normally incident beam into an ON ($+ 24^\circ$ optical deflection) and OFF  ($- 24^\circ$ optical deflection) wavefront. These wavefronts are subsequently recombined by re-imaging them onto the very same DMD chip. Because the OFF part of the wavefront is utilized, but delayed by the equivalent of a  $\pi$-phase shift, an arbitrary binary phase pattern can be imprinted, while the spatial dispersion is auto-corrected. This enables the use of the DMD chip for binary phase modulation of $200\times 200$ pixels at 20~kHz with pulsed laser sources, which is demonstrated by using the device in two-photon excitation fluorescence applications.

\begin{figure}[!htb]
\centering\includegraphics[width=\linewidth]{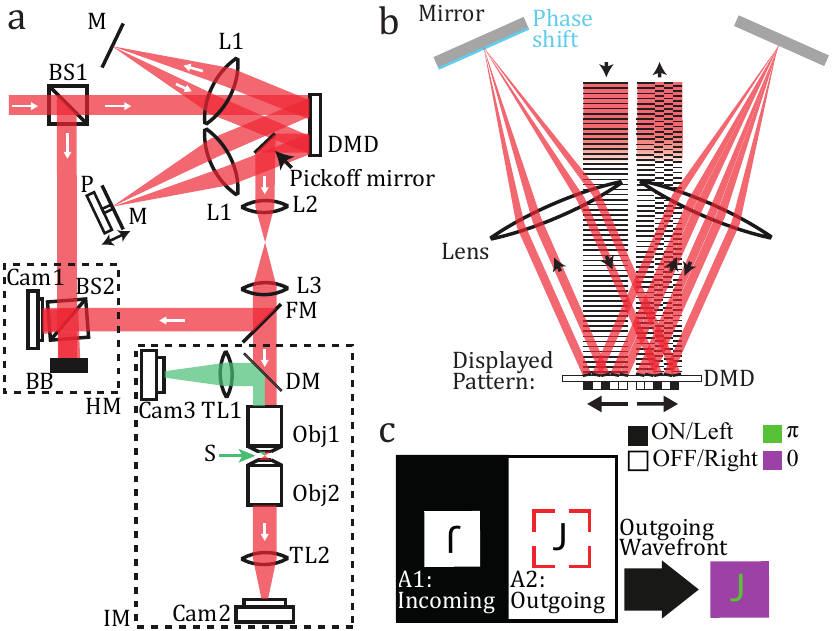}
\caption{  Optical setup and principle of operation: (a) Optical setup with the binary phase modulator, the off-axis holography module (HM) and the imaging module (IM). (b) Operating principle of the binary phase modulator: The incoming beam is split into two wavefronts according to the displayed DMD pattern. It is re-imaged onto the same DMD by a lens-mirror combination. If the wavefront is incident at the same mirror distribution of the first pass, it recombines and propagates outwards and parallel to the incident beam. A phase shift can be imprinted by delaying one wavefront with respect to the other, e.g. phase mask, or lens displacement. (c) Only a subset of the DMD was used as an active area. The beam was incident in area A1 and by the two imaging systems onto area A2. In A2 the pattern of A1 rotated by $180^\circ$ was displayed to recombine the wavefronts. Mirrors outside of the two active areas were flipped into the opposite direction, so that all light outside of area A1 was reflected out of the beampath, which resulted in a square outgoing wavefront. Since one of the arms had a path difference equal to a $\pi$-phase shift, the outgoing wavefront was modulated as desired. 
}
\label{fig:Fig1}
\end{figure}

Central to our binary phase modulator is a digital micromirror device (DMD, V-7001, $1024\times768$ mirrors, Vialux). Each pixel of this DMD is a small square mirror that can be tilted to  $\pm 12^\circ$ along its diagonal by setting it into either of two states, conventionally labeled "ON" or "OFF" (although we are utilizing both the ON and OFF state, rather than blocking light from OFF-state mirrors, we will adhere to this nomenclature). As a whole, the DMD modulates coherent light similar to a blazed grating and thus can only be used efficiently for certain combinations of wavelength and incident angle. Since our application required a $0^\circ$ incident angle with respect to the DMD window normal, the used wavelengths had to fulfill the grating equation $\sqrt{2}d\tan 12^\circ=m\lambda$ . Here $ d=13.68~$\textmu m is the DMD pixel pitch,  $\lambda$ is the (central) wavelength of the light used and $m$ is a positive integer.  We used a Ti:Sapphire laser  (Mai Tai HP, Spectra Physics) with a central wavelength of 820 nm to fulfill the above condition. The beam was split before the DMD into a reference and a probe beam using a 50:50 non-polarizing beamsplitter (BS1), Fig. 1(a). The probe beam was normally incident onto one section of the DMD, which was mounted at $45^\circ$ to ensure, that the tilt axis of the mirrors were perpendicular to the table. Therefore the optical axis of the system was at all times parallel to the optical table .
The beam was positioned at the center of a designated active area A1, which was slightly offset from the geometric center of the DMD array. The beam slightly over-filled the $200\times 200$ pixel active area of A1 (2.192 mm x 2.192 mm), Fig.~\ref{fig:Fig1}(a).
The DMD split the beam within the area A1 into two wavefronts according to the pattern displayed, with each beam propagating in the $\pm 2\times 12^\circ = \pm 24^\circ$ directions, Fig.~\ref{fig:Fig1}(b).
Along both directions, we placed two imaging arms mounted on a translation stage. Each imaging arm consisted of a lens (L1:  $d=2^{\prime\prime}$, $f_1=150$~mm) on a xy-translation mount and a mirror located at the back focal plane of the lens. In one arm, the mirror was attached to a linear piezo actuator (P: S-310.10, Physik Instrumente), which allowed for mirror displacement with nanometer precision. The front focal plane and optical axis of the two lenses were then aligned to intersect the center of the DMD sensor by translating the lens mounts and the whole imaging arm.
In this configuration, both arms acted as independent 4f imaging systems. Although the DMD surface is tilted with respect to the imaging planes by $\pm 24^\circ$, the 1x magnification imaging systems provide a mapping of the optical field within area A1 back to the surface of the DMD\cite{Botcherby:2008bk}. At each micromirror within area A1, we have a means to select whether the incident light goes through the left or the right imaging arm. We will introduce a phase delay in the left imaging arm to achieve phase modulation.

The light from each imaging arm then returns to a new DMD subarea A2, which is located on the opposite side of the center of the DMD with respect to subarea A1, Fig.~\ref{fig:Fig1}(c). If the patterns displayed on both areas A1 and A2 are identical but rotated by $180^\circ$ with respect to one another, the wavefronts from both arms recombined by reflecting at subarea A2 and left the device parallel to the incident wavefront, Fig.~\ref{fig:Fig1}(c).  Importantly, the second-pass of the optical field from each imaging arm over the DMD reverses all spatial dispersion caused by the first interaction of light with the DMD surface. This allows us to reconstitute the original wavefront with minimal dispersion (apart from dispersion caused by the imprinted phase pattern and the optical elements in the path).
Next, we apply a $\pi$-phase shift to the wavefront within the OFF-imaging arm by changing its optical path length via the piezo actuated mirror and the translation stage. This imprints a binary phase pattern onto the recombined wavefront within area A2, Fig.~\ref{fig:Fig1}(b,c). The wavefront reflected off area A2 is then redirected with a $1^{\prime\prime}$ D-shaped mirror, to spatially separate it from the incoming beam. This binary phase modulated field may then enter any imaging or projection system that requires phase-only wavefront shaping control. 

To test our device, the modulated beam at the surface of the DMD was relayed by a telescope  (L2: d=$2^{\prime\prime}$, $f=200$~mm L3: $f=250$~mm) into an off-axis holography module (HM, Fig.~\ref{fig:Fig1}(a)) and an imaging module (IM, Fig.~\ref{fig:Fig1}(a)). We could switch between the two with a flip mirror (FM).
In the holography module, the wavefront at the surface of the DMD was imaged onto a camera (Cam1, acA2440-75um, Basler ace). There it was made to interfere with the path-matched reference beam by using a second 50:50 non-polarizing beamsplitter (BS2)  in an off-axis holography configuration \cite{Leith:1962ep}, where the excess light was absorbed by a beam block (BB). The resulting interference pattern allowed us to assess the pixel-to-pixel alignment, the pulse coherence and the phase of the outgoing wavefront.

In the imaging module, we directed the tailored wavefront onto the back focal aperture of a microscope objective (Obj1: $f=4.6$~mm, $\text{NA}=0.66$, $\text{WD}=0.49$~mm, L-40x, Newport) using an effective numerical aperture (NA) of 0.4. Doing so allowed us to probe the transformation of the DMD surface at the image plane of Obj1. We probed the image plane both in transmission with a separate imaging system as well as in reflection in an epi-fluorescence imaging geometry. 

For the transmission imaging experiment, an identical objective (Obj2) was placed face to face to the first objective in a 4f configuration. The image plane located between the objective lenses was then imaged onto a camera (Cam2: BFS-U3-51S5M-C, FLIR) via a tube lens (TL2: $f=200$~mm).
Additionally we placed a dichroic mirror (DM) before the microscope objective (Obj1) and imaged all of the returning fluorescent light from a fluorescent sample (S) placed in front of Obj1 onto camera Cam3 (BFS-U3-51S5M-C, FLIR, TL1: $f=200\text{~mm}$). We used a fluorescein water solution as the fluorescent sample in all of the conducted experiments, which had a sufficiently high two photon excitation cross section at 820~nm  \cite{XU:2003to}. Finally, whenever synchronized control of the DMD and the cameras was needed, both were triggered via a multifunction I/O device (USB-6363, National Instruments).

\begin{figure}[htbp]
\centering\includegraphics[width=\linewidth]{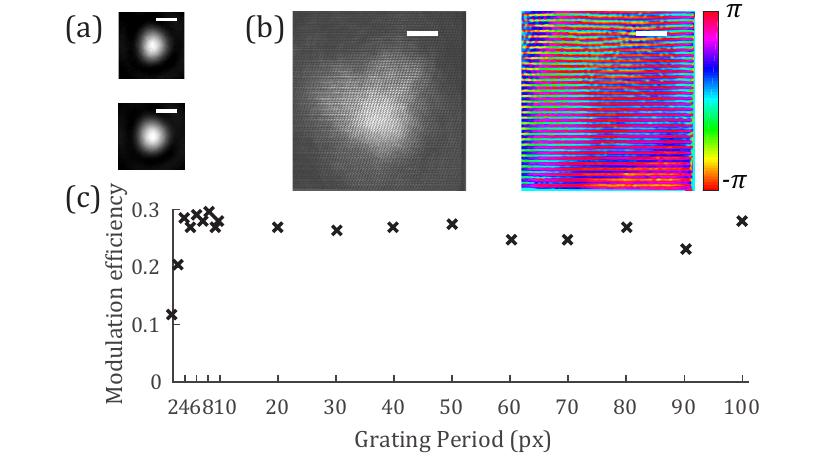}
\caption{ Auto-correction of spatial dispersion and diffraction efficiency of the DMD-based binary phase modulator. (a) Beam profile in continuous wave mode (top) and pulsed mode (bottom) have identical beam waists, scale bars 1~\textmu m. (b)  Off-axis hologram of the surface of the DMD (Area A2)  displaying a binary phase grating used for diffraction efficiency measurement with period 6~px ($200\text{~px} \times 200\text{~px}$) (left) and the corresponding reconstructed phase distribution (right), scale bars $500$~\textmu m. (c) Plot of the diffraction efficiencies of binary phase gratings displayed by the DMD as a function of grating period.
}
\label{fig:Fig2}
\end{figure}

To characterize our device, we first measured its overall transmission efficiency by turning all mirrors to the ON position. We then measured the output power after the pick-off mirror to be 25\%  of the incident power. 
To assess whether our optical setup properly canceled spatial dispersion by the DMD, we imaged the focus of the laser beam in the imaging module when running the laser both in continuous wave (monochromatic) as well as in pulsed mode. A comparison of the two resulting images,~Fig.~\ref{fig:Fig2}(a), indicates the absence of severe spatial dispersion for the pulsed beam.

We then characterized the modulation efficiency of the DMD by projecting binary phase gratings with various periods. We first verified that the desired phase modulation was imprinted on the probe beam. This was achieved by reconstructing the amplitude and phase of the wavefront at the DMD surface from the off-axis hologram recorded with the holography module, Fig.~\ref{fig:Fig2}(b).
Each binary phase grating generated by the DMD and projected to the back aperture of objective Obj1, will produce a number of diffraction spots at the imaging plane of the imaging module. The modulation efficiency of each binary phase grating was calculated as the ratio between the total intensity within the 0th order when all of the DMD pixels were set to the ON position (flat phase wavefront), and the total intensity measured for the first diffraction order after displaying a grating. We plot this ratio as a function of grating period in Fig.~\ref{fig:Fig2}(c). Here, we see that the device had a minimum modulation efficiency of 10\% at small grating periods, while it converged to an average of 27\% for periods larger than 3 pixels.
Combining this diffraction efficiency with the 25\% all-ON efficiency results in a total device efficiency of 7\%, Fig.~\ref{fig:Fig2}(c).

\begin{figure}[htbp]
\centering\includegraphics[width=\linewidth]{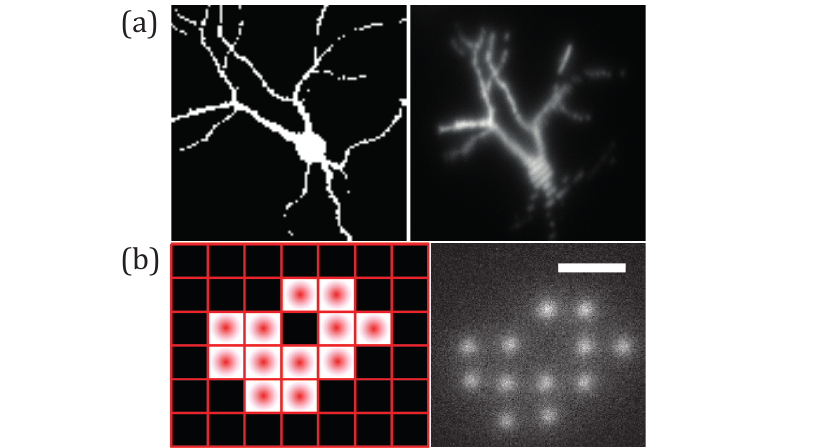}
\caption{  High-speed point scanning and two-photon fluorescence excitation: (a) Left: $100\times 100$ pixels pictogram of a neuron with 790 illuminated pixels. Right: single image of two-photon excited fluorescence of the point scanned pictogram (point rate 20~kHz), exposure: $790 \times 50\text{~\textmu s} =39.5\text{~ms}$, scale bar $10 \text{~\textmu m}$. (b) Left: Game of Life grid with LWSS glider figure (12 px); Right:  the same arena as point scanned two-photon fluorescence image, exposure: $600\text{ ~\textmu s}$, scale bar $5 \text{~\textmu m}$, the periods of the displayed gratings were between 3.3 px and 20 px. (see \textbf{Visualization 1})
}
\label{fig:Fig3}
\end{figure}

To illustrate potential applications as well as to further test our device, we performed several proof-of-principle two-photon excitation experiments. We used the phase-modulator to scan a laser focus over a fluorescent sample of interest by displaying binary phase gratings similar to Fig.~\ref{fig:Fig2}(b), which are the binary equivalent of a phase ramp.
The theoretically maximum point displacement from the 0th order of our setup was given by $\tan^{-1}\left(\frac{\lambda}{2 M d}\right) f_{\mathrm{obj}}=110$~\textmu m. Here $\lambda$ is the wavelength, $M=1.25$ is the magnification from the DMD surface to the back focal aperture of the objective, $f_{\mathrm{obj}}$ is the focal length of the objective and $d$ is the pixel pitch of the DMD. Moreover since the Fourier transform of a binary phase pattern is inversion symmetric, the maximum number of  independently controlled modes within a square field is $100\times 100$.
More precisely, we raster-scanned a pictogram of a neuron consisting of 790 points within an area of $55\text{~\textmu m}\times 55\text{~\textmu m}$, offset from the 0th order by $15 $~\textmu m on a fluorescence target by successively displaying 790 different binary phase gratings . The resulting two-photon fluorescence excited spots were recorded by the epi-fluorescence detection system of the imaging module. Scanning at 20 kHz allowed us to complete the entire 790-point diagram in 39.5~ms, Fig.~\ref{fig:Fig3}(a).
To further illustrate the high speed of our binary phase modulation we created and projected a point-scanned movie of the lightweight spaceship (LWSS) of Conway's game of life onto the fluorescent sample,  Fig.~\ref{fig:Fig3}(b).
In this experiment, we moved the glider through a $25\text{~\textmu m} \times 25\text{~\textmu m}$ area. Every camera frame contained at a maximum 12 points and thus required displaying a maximum of 12 different gratings.  Scanning for each movie frame at 20~kHz thus required $12\times 50  \text{~\textmu s} = 600  \text{~\textmu s}$. (Fig.~\ref{fig:Fig3}(b) and Visualization 1.

In this paper we have introduced a fast binary phase modulator design based on a digital micromirror device (DMD) with auto-cancelation of spatial dispersion that does not require a separately aligned dispersion correction element. We characterized the efficiency of our device and verified the absence of spatial dispersion within the light it modulates. To illustrate both its speed and potential application within microscopy, we used the device for point-scanning of a pulsed laser beam over a fluorescent target and imaged the excited two-photon fluorescence with a conventional camera. The device allowed for light modulation at a rate of 20~kHz as demonstrated in the supplementary movie.
The theoretical efficiency of our modulation strategy is 40.5\%~\cite{Goodman:2005tc} in comparison to the 7\% achieved empirically, which is a combined effect from optical losses and a suboptimal modulation efficiency of the displayed grating. 

The efficiency without any phase modulation was 25\%. This largely results from losses at the window and the aluminum micromirrors of the DMD. The single pass transmittance specification of our DMD window is given as $\approx 90\%$ at 820~nm. This implies that the overall efficiency of the device can increase by 20\% if the window transmittance of the DMD were to be 96\%, a value typical for AR coatings in the NIR range. By using the DLP4500NIR (Texas Instruments), instead of the DLP7500 used here, one could trade off speed for efficiency, since the maximum switching rate of the DLP4500NIR (that includes this coating) is given at around 4~kHz. In addition, this alternative DMD has smaller pixels and would therefore need to operate at a different wavelength, for which the blazing condition is again satisfied.
Alternatively, for lasers with a low repetition rate the blazing condition could be dynamically modified by strobing the DMD during a mirror flip as described in \cite{Smith:2017ba}.

The measured modulation efficiency of the displayed binary phase grating was below the theoretically possible modulation efficiency of $40.5\%$. It varied from approx. 10\% for high frequency gratings to approximately 27\% for gratings with a period above 3~pixel. This discrepancy and trend is most consistent with an effective crosstalk between the DMD pixels possibly caused by imperfect re-imaging of the DMD surface onto itself and losses of light at edges between ON and OFF areas. Constructing better re-imaging arms, perhaps a dedicated optical design for the tilted planes, may help address this latter issue. Additionally, an active alignment system, which would control the degrees of freedom of the DMD and the two imaging lenses (L1), while monitoring the phase profile of the DMD surface might further improve the quality of the optical re-imaging.

Further improvement could be made by increasing the number of modes controlled by our device. Currently the active area of $200\times200$ pixels is limited by aberrations within the two imaging arms, which limit the quality of both re-imaged wavefronts at the DMD surface and manifest in the non-uniform phase steps in Fig.~\ref{fig:Fig2}(b). 
Using the full active DMD surface would require us to image an area of approx. $13.8\text{~mm} \times 10.3\text{~mm}$, which is tilted with respect to the focal plane onto the imaging system, at a pixel-level accuracy. The maximal distance of the DMD surface from the focal plane would then be about $1.5\text{~mm}$, a distance which can theoretically be imaged without introducing spherical aberrations given the moderately low NA needed.~\cite{Botcherby:2008bk}
If achieved by an improved optical design, this would allow us to control the binary phase of $512\times 768$ modes.

In previous studies, a DMD was used as a multi-level phase modulator for a continuous wave laser at an efficiency of 7\% by using a Lee hologram \cite{DonConkey:2012vd}. Recently the Lee hologram method was also established for pulsed light by using an additional grating for dispersion correction with a reported efficiency of 4\% \cite{Geng:2017km}. While our device only implements a binary phase modulation, the gain in efficiency, albeit small, can be a decisive factor in an optical setup with many optical components.

The greatest advantage of the proposed device is its high update rate, which is only rivaled by deformable mirrors (DMs) or ferroelectric spatial light modulators. Although the efficiency of DMs is very high, the purchasing price of the DMs is approximately 3 to 5-fold higher than of a DMD and the actuator count is presently limited to around 4000. Ferroelectric spatial light modulators, on the other hand, offer higher pixel count at an efficiency of around 5\%, but need to display the inverse version of every displayed phase pattern immediately after, which introduces 50\% dead time for most applications and limits the speed to the lower kHz range. In comparison  the speed and pixelcount of the DMD-based device make it a cost effective candidate for optogenetic stimulation or fast multiplexed aberration correction in two-photon point-scanning microscopy.

\paragraph*{}
The work was supported by European Research Council (ERC-2016-StG-714560),  Einstein Foundation Berlin, the DFG (EXC 257 NeuroCure), HFSP,  and Krupp Foundation.
\paragraph*{}
The authors thank R. Horstmeyer and E. Bobrov for critically reviewing the manuscript.

\bibliography{bib}
\bibliographystyle{abbrv}

\end{document}